\documentclass[aps,prb,reprint,showpacs]{revtex4-1}

\usepackage{graphicx}
\begin{document}

\title{Ferroelectric phase transition in orthorhombic CdTiO$_3$: First-principles
studies}


\author{Alexander I. Lebedev}
\email[]{swan@scon155.phys.msu.ru}
\affiliation{Physics Department, Moscow State University, Moscow, 119991 Russia}

\date{\today}

\begin{abstract}
The crystal structures and phonon spectra of orthorhombic cadmium titanate with
the $Pbnm$ structure and of its two possible ferroelectrically distorted
phases with $Pbn2_1$ and $Pb2_1m$ space groups were calculated from first
principles within the density functional theory. The obtained structural
parameters and frequencies of Raman- and infrared-active modes are in good
agreement with available experimental data for the $Pbnm$ phase. Expansion of
the total energy in a Taylor series of two order parameters showed that the
ground state of the system corresponds to the $Pbn2_1$ structure into which
the $Pbnm$ phase transforms through a second-order phase transition without
intermediate phases. A substantial discrepancy between the calculated and
experimentally observed lattice distortions and spontaneous polarization in
the polar phase was explained by quantum fluctuations as well as by existence
of twinning and competing long-period structures.

\texttt{DOI: 10.1134/S1063783409040283}
\end{abstract}


\maketitle

\section{Introduction}

The calculations of phonon spectra of ten titanates \emph{A}TiO$_3$
(\emph{A} = Ca, Sr, Ba, Ra, Cd, Zn, Mg, Ge, Sn, Pb) with the perovskite
structure in Ref.~\onlinecite{PhysSolidState.51.362} have revealed that
orthorhombic phases of CdTiO$_3$, ZnTiO$_3$, and MgTiO$_3$ with the
$Pbnm$ structure exhibit the ferroelectric instability which appears
in their phonon spectra as one or two unstable modes with symmetries
$B_{1u}$ and $B_{2u}$ at the $\Gamma$~point. In this work, the
characteristics of these modes are considered and the results of
first-principles calculations of the crystal structure, phonon spectra, and
spontaneous polarization of the parent orthorhombic CdTiO$_3$ phase with
$Pbnm$ structure and its two ferroelectrically distorted orthorhombic
modifications (with $Pbn2_1$ and $Pb2_1m$ space groups) are compared with
available experimental data.

The ferroelectric phase transition in cadmium titanate was discovered by
Smolenskii~\cite{DAN.70.405} in 1950. Since then, this phase transition was
studied by dielectric~\cite{Ferroelectrics.217.137,Ferroelectrics.259.85,
PhysSolidState.43.2146,Ferroelectrics.270.381} and
X-ray~\cite{Ferroelectrics.217.137,Ferroelectrics.259.85,
Ferroelectrics.284.107} methods; using Raman
scattering;~\cite{Ferroelectrics.217.137,PhysSolidState.47.337}
IR and submillimeter reflectance spectroscopy;~\cite{PhysSolidState.47.547}
pyroelectric current measurements.~\cite{Ferroelectrics.259.85}  The specific
features of the phase transition in CdTiO$_3$ are a smallness of the lattice
distortions and a considerable scatter in the data on the phase transition
temperature (50--82~K) and spontaneous polarization (0.002--0.009~C/m$^2$).
Even the data on the structure of the low-symmetry phase are contradictory. In
particular, the study of the dielectric properties~\cite{Ferroelectrics.270.381}
suggests that the $2_1/m$ axis becomes polar, whereas structural studies using
synchrotron radiation suggest that it is the $2_1/n$ axis that becomes
polar.~\cite{Ferroelectrics.284.107}  In Refs.~\onlinecite{PhysSolidState.43.2146,
PhysSolidState.47.337,PhysSolidState.47.547}, the analysis of the temperature
dependence of the dielectric constant and the Raman lines intensity allowed
the authors to suggest that upon cooling one more phase transition accompanied
by a change in the polarization direction occurs in CdTiO$_3$ at about 50~K.

The paper by Fabricius and L{\'o}pez Garc{\'i}a~\cite{PhysRevB.66.233106} is
the only work in which CdTiO$_3$ was studied from first principles. In this
work, the authors calculated the electric field gradient at cadmium atoms for
different sets of atomic coordinates proposed in the literature for $Pbnm$
and $Pbn2_1$ structures and revealed that the latter structure transforms into
a more stable nonpolar $Pbnm$ phase during relaxation.

In view of the discrepancies between the experimental data available in the
literature for cadmium titanate, it is expedient to perform first-principles
calculations in order to elucidate the causes responsible for these
contradictions and to resolve them.

\section{Calculation technique}

The calculations were performed within the density functional theory with the
pseudopotentials and the plane-wave expansion of wave functions as implemented
in the \texttt{ABINIT} code.~\cite{abinit}  The exchange-correlation interaction
was described in the local-density approximation (LDA).~\cite{PhysRevB.23.5048}
As pseudopotentials, we used the optimized
separable nonlocal pseudopotentials,~\cite{PhysRevB.41.1227}  generated with
the \texttt{OPIUM} code, to which the local potential was added in order to
improve their transferability.~\cite{PhysRevB.59.12471}  The parameters used
for constructing the pseudopotentials, the results of their testing, and other
details of calculations are described in Ref.~\onlinecite{PhysSolidState.51.362}.

\section{Results of the calculations}

\subsection{Phonon spectrum of the $Pbnm$ phase}

\begin{table}
\caption{\label{table1}Calculated frequencies (in cm$^{-1}$) of the softest
ferroelectric modes in four titanates with the orthorhombic $Pbnm$ structure
and tetragonal strontium titanate.}
\begin{ruledtabular}
\begin{tabular}{cccccc}
Mode & SrTiO$_3$ & CaTiO$_3$ & CdTiO$_3$ & ZnTiO$_3$ & MgTiO$_3$ \\
\hline
$A_{2u}$ & 55  & --- & ---    & ---    & ---    \\
$E_u$    & 39  & --- & ---    & ---    & ---    \\
$B_{3u}$ & --- & 82  & 54     & 73     & 115    \\
$B_{2u}$ & --- & 97  & 81$i$  & 54$i$  & 81     \\
$B_{1u}$ & --- & 82  & 104$i$ & 103$i$ & 133$i$ \\
\hline
$\Gamma_{15}$ & 68$i$ & 165$i$ & 187$i$ & 240$i$ & 260$i$ \\
\end{tabular}
\end{ruledtabular}
\noindent\raggedright
{\footnotesize Note: The frequencies of unstable TO phonons at the
$\Gamma$~point of the Brillouin zone in the cubic parent phases are shown
in the lowest row.~\cite{PhysSolidState.51.362} }
\end{table}

The calculated frequencies of the softest phonons at the $\Gamma$~point, which
are responsible for the ferroelectric instability in orthorhombic CaTiO$_3$,
CdTiO$_3$, ZnTiO$_3$, and MgTiO$_3$ compounds with the $Pbnm$ structure and
in tetragonal SrTiO$_3$ with the $I4/mcm$ structure, are given in Table~\ref{table1}.
In CaTiO$_3$ and SrTiO$_3$, all phonons are stable (this corresponds
to experiment), whereas in the other three titanates one or two unstable modes
(whose frequencies are imaginary) arise. The strongest instability
in these compounds is associated with the $B_{1u}$ phonon, which can induce
the $Pbnm \to Pbn2_1$ phase transition. In CdTiO$_3$ and ZnTiO$_3$, the
phonon with the $B_{2u}$ symmetry is also unstable. This phonon can induce
the $Pbnm \to Pb2_1m$ phase transition. It should be noted that, among the
three compounds under consideration, the $Pbnm$ structure can be obtained only
for CdTiO$_3$ (ZnTiO$_3$ and MgTiO$_3$ usually crystallize in the ilmenite
structure). This is why hereafter only the properties of cadmium titanate will
be considered.

\begin{table}
\caption{\label{table2}Lattice parameters $a$, $b$, and $c$ (in {\AA}) and
atomic coordinates for CdTiO$_3$ with the $Pbnm$ structure.}
\begin{ruledtabular}
\begin{tabular}{cccc}
Parameter & This work & \multicolumn{2}{c}{Experiment} \\
          &           &  Ref.~\onlinecite{ActaCrystC.43.1668} & Ref.~\onlinecite{Ferroelectrics.284.107}$^*$ \\
\hline
$a$    & 5.2427     & 5.3053     & 5.284 \\
$b$    & 5.3815     & 5.4215     & 5.403 \\
$c$    & 7.5744     & 7.6176     & 7.590 \\
Cd$_x$ & $-$0.01017 & $-$0.00847 & $-$0.00891 \\
Cd$_y$ & +0.04637   & +0.03873   & +0.03997 \\
Cd$_z$ & +0.25000   & +0.25000   & +0.25000 \\
Ti$_x$ & +0.00000   & +0.00000   & +0.00000 \\
Ti$_y$ & +0.50000   & +0.50000   & +0.50000 \\
Ti$_z$ & +0.00000   & +0.00000   & +0.00000 \\
O1$_x$ & +0.10130   & +0.0902    & +0.0918 \\
O1$_y$ & +0.46252   & +0.4722    & +0.4714 \\
O1$_z$ & +0.25000   & +0.25000   & +0.2500 \\
O2$_x$ & +0.69348   & +0.7008    & +0.70083 \\
O2$_y$ & +0.30304   & +0.2969    & +0.29660 \\
O2$_z$ & +0.05341   & +0.0472    & +0.04783 \\
\end{tabular}
\end{ruledtabular}
\noindent{\footnotesize $^*$ At $T = 150$~K. \hfill}
\end{table}

The calculated lattice parameters and equilibrium atomic coordinates for
orthorhombic CdTiO$_3$ with the $Pbnm$ structure are compared with available
experimental data in Table~\ref{table2}. It is seen that the results of the
calculations are in good agreement with the experimental data. It should be
noted that a better agreement is observed for the experimental data obtained
at lower temperatures. A small systematic underestimation of the calculated
lattice parameters is typical of the LDA approximation used in this work.

The vibrational spectrum of a crystal with the $Pbnm$ space group consists
of 60~modes, including 24~Raman-active modes (with symmetries $A_g$, $B_{1g}$,
$B_{2g}$, and $B_{3g}$), 25 IR-active optical modes (with symmetries $B_{1u}$,
$B_{2u}$, and $B_{3u}$), 8 so-called silent $A_u$ optical modes, and three
acoustic modes inactive in the optical spectra.

\begin{table*}
\caption{\label{table3}Frequencies $\nu_i$ of the optical modes active in Raman
and IR reflectance spectra and oscillator strengths $f_i$ for IR-active modes
in CdTiO$_3$ with the $Pbnm$ structure.}
\begin{ruledtabular}
\begin{tabular}{cccccccc}
Mode  & \multicolumn{3}{c}{$\nu_i$ (cm$^{-1}$)} & Mode & $f_i \cdot 10^3$, & \multicolumn{2}{c}{$\nu_i$ (cm$^{-1}$)} \\
      & This work & Exp. (Ref.~\onlinecite{Ferroelectrics.217.137}) & Exp. (Ref.~\onlinecite{PhysSolidState.47.337}) & & arb. units & This work & Exp. (Ref.~\onlinecite{PhysSolidState.47.547}) \\
\hline
$A_g$ & 96        & 95   & 99  & $B_{3u}$ & 3.87 & 54  & 61? \\
      & 128       & 123  & 125 &          & 1.27 & 104 & 111 \\
      & 195       & 190  & 194 &          & 0.60 & 167 & 165 \\
      & 295       & 295  & 299 &          & 0.38 & 275 & 284 \\
      & 414       & 390  & 390 &          & 0.30 & 302 & 306 \\
      & 449       & 461  & 465 &          & 0.71 & 379 & 383 \\
      & 512       & ---  & 496 &          & 0.13 & 423 & --- \\
$B_{3g}$ & 140    & 135,141 & 144 &       & 0.17 & 445 & 458 \\
         & 211    & ---  & --- &          & 0.96 & 513 & 525 \\
         & 353    & 342  & 346 & $B_{2u}$ & 5.98 & 81$i$ & 61** \\
         & 486    & ---  & 479* &         & 0.62 & 90  & 111** \\
         & 683    & ---  & ---  &         & 0.86 & 177 & 177 \\
$B_{2g}$ & 117    & ---  & 114  &         & 0.01 & 202 & --- \\
         & 280    & ---  & 307? &         & 0.37 & 321 & 321 \\
         & 445    & ---  & 459* &         & 0.14 & 345 & 349 \\
         & 488    & ---  & 509* &         & 0.12 & 437 & --- \\
         & 764    & ---  & ---  &         & 1.04 & 476 & 483 \\
$B_{1g}$ & 111    & 110  & 115  &         & 0.02 & 530 & 525 \\
         & 142    & 141  & 141  & $B_{1u}$ & 3.42 & 104$i$ & 44** \\
         & 205    & ---  & ---  &          & 0.89 & 60 & 96** \\
         & 356    & ---  & ---  &          & 0.78 & 128 & 143** \\
         & 448    & ---  & ---  &          & 1.24 & 224 & 225 \\
         & 492    & ---  & 504  &          & 0.62 & 396 & 387 \\
         & 738    & ---  & ---  &          & 0.02 & 428 & --- \\
         &        &      &      &          & 1.19 & 481 & 511 \\
\end{tabular}
\end{ruledtabular}
\noindent{\footnotesize $^*$ Components of a broad weakly structured line. \hfill}

\noindent{\footnotesize $^{**}$ A strong deviation from the calculations is a result of anharmonicity. \hfill}
\end{table*}

The calculated frequencies of modes active in IR reflectance and Raman
scattering are compared with available experimental
data~\cite{Ferroelectrics.217.137,PhysSolidState.47.337,PhysSolidState.47.547}
in Table~\ref{table3}. For the Raman-active modes, the results of the
calculations agree well with the experimental data obtained on ceramic
samples~\cite{Ferroelectrics.217.137} and single
crystals;~\cite{PhysSolidState.47.337}  the typical relative deviation of
frequencies is about 3\%. A comparison of the mode frequencies determined
from polarized Raman scattering spectra~\cite{PhysSolidState.47.337} with the
results of our calculations shows that the peaks observed at 303 and
392~cm$^{-1}$ in the $yy$ polarization (in the crystal setting accepted in
Ref.~\onlinecite{PhysSolidState.47.337}) is most likely erroneously assigned
to the $B_{1g}$ modes (in our crystal setting). The positions of these peaks
are close to the positions of the peaks of the $A_g$ modes, which can also
be observed in the above polarization. The experimental peak at 307 cm$^{-1}$
in the $yz$ polarization, assigned to the $B_{2g}$ or $B_{3g}$ modes, is
most likely a ``leak'' of the $A_g$ mode.

A comparison of the calculated mode frequencies with the results obtained
from an analysis of IR reflectance spectra~\cite{PhysSolidState.47.547}
appeared to be a more complex problem. A direct comparison of the experimental
frequencies with the frequencies calculated for the symmetries of modes indicated
in Ref.~\onlinecite{PhysSolidState.47.547} revealed their significant disagreement.
In order to reliably identify the experimentally observed modes, the oscillator
strength $f_i$ was additionally calculated for each mode (the oscillator
strength characterizes the contribution of the mode into the complex dielectric
constant) and it was supposed that in Ref.~\onlinecite{PhysSolidState.47.547}
the symmetries of the observed modes were identified incorrectly.
This can be a result of the high twin density characteristic of CdTiO$_3$
crystals,~\cite{PhysSolidState.47.337,ActaCrystC.43.1668,PhysSolidState.42.1329}
which can undoubtedly manifest itself in measurements on samples with a large
area.

\begin{figure}
\includegraphics[scale=0.75]{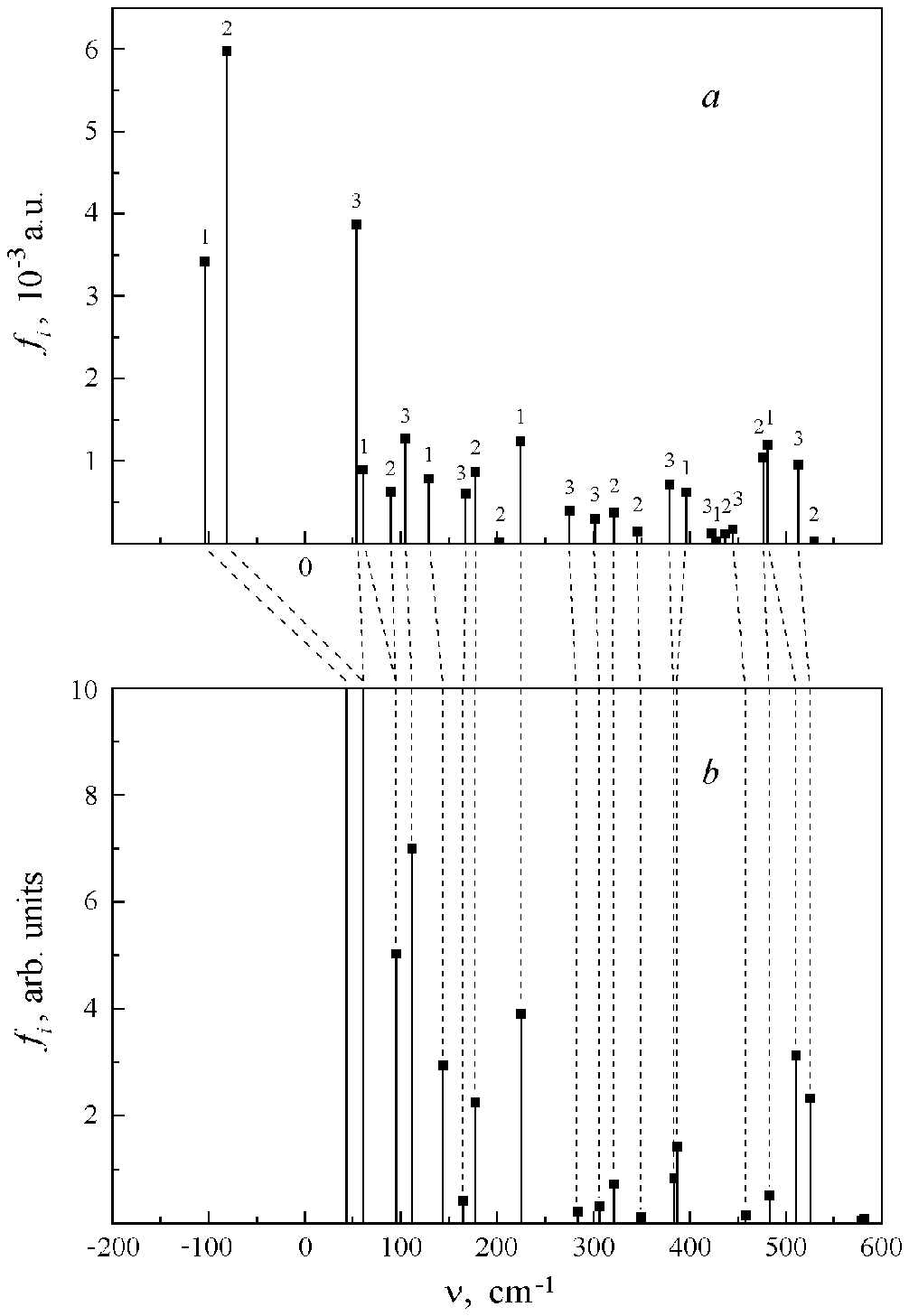}
\caption{Comparison of (a) the calculated frequencies and oscillator strengths
of the IR-active modes for cadmium titanate with the $Pbnm$ structure with
(b) the corresponding parameters obtained from an analysis of the IR reflectance
spectra. Labels near the points indicate the mode symmetry ($B_{1u}$, $B_{2u}$,
$B_{3u}$).}
\label{fig1}
\end{figure}

The calculated frequencies and oscillator strengths for the CdTiO$_3$
crystal with the $Pbnm$ structure are compared with the results of
the IR reflectance spectra studies in Fig.~\ref{fig1}. Under the assumption
that the modes identified as $B_{3u}$ in Ref.~\onlinecite{PhysSolidState.47.547}
for the $Pnma$ crystal setting correspond to the $B_{1u}$ modes in our $Pbnm$
setting and that the $B_{1u}$ modes in Ref.~\onlinecite{PhysSolidState.47.547}
correspond to the $B_{2u}$ and $B_{3u}$ modes in our setting, the agreement
between the calculated and experimental data becomes more reasonable
(Fig.~\ref{fig1}, Table~\ref{table3}). A considerable shift in frequencies of
three softest $B_{1u}$ modes and two softest $B_{2u}$ modes
as compared to the calculated frequencies is explained by
anharmonicity. The modes with calculated frequencies of 54 and 104~cm$^{-1}$
are most likely indistinguishable in the experiment from the other modes, and
four modes with the smallest oscillator strengths are not observed in the spectra
at all.

\subsection{Lattice distortions accompanying the ferroelectric phase transitions}
\label{Sec3.2}

\begin{figure*}
\includegraphics[scale=0.75]{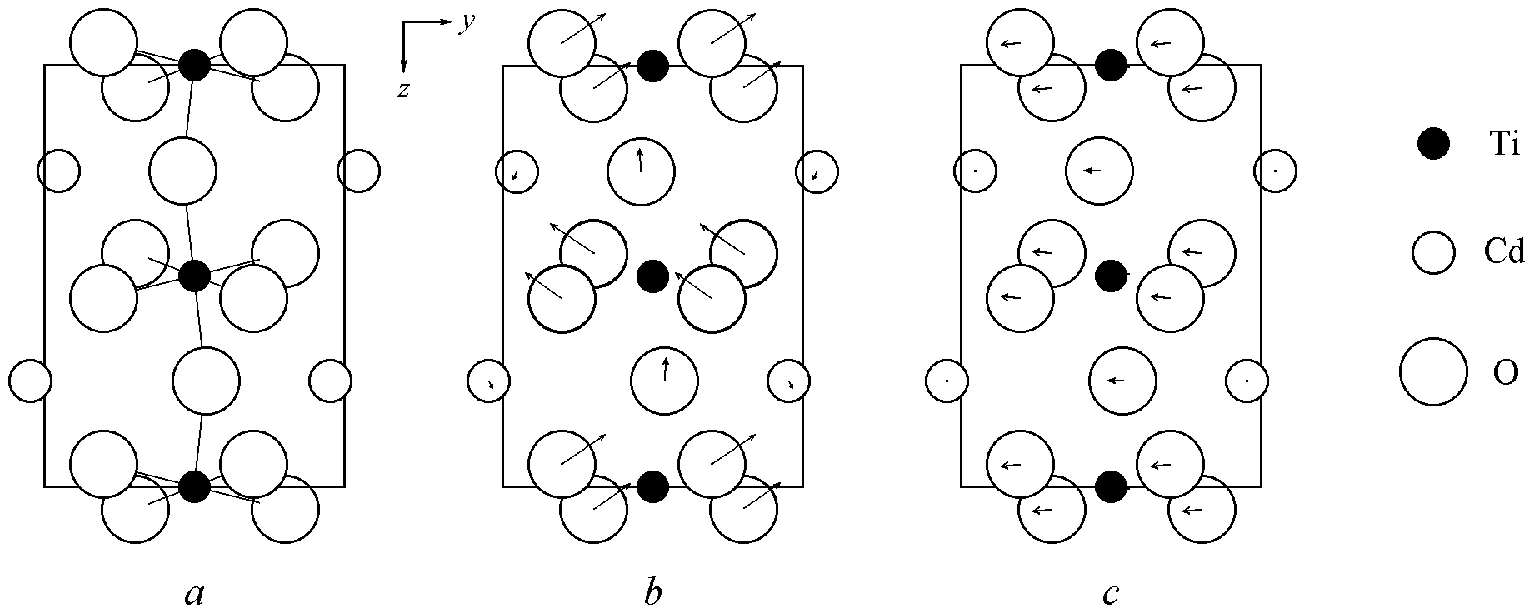}
\caption{(a) Projection of the structure of the orthorhombic CdTiO$_3$ phase
with the $Pbnm$ structure onto the $bc$ plane and (b, c) the character
of atomic displacements accompanying the ferroelectric transitions to the
(b) $Pbn2_1$ and (c) $Pb2_1m$ phases.}
\label{fig2}
\end{figure*}

As shown above, the used technique of first-principles calculations describes
well the properties of orthorhombic CdTiO$_3$ with the $Pbnm$ structure.
Now, we consider the properties of ferroelectrically distorted phases
of cadmium titanate. The equilibrium atomic positions in the distorted
phases were determined by relaxing the Hellmann--Feynman forces in
structures obtained from the parent nonpolar $Pbnm$ phase by adding a
perturbation with the $B_{1u}$ symmetry (for the $Pbn2_1$ phase) or $B_{2u}$
symmetry (for the $Pb2_1m$ phase). The structure of the nonpolar $Pbnm$ phase
and the character of atomic displacements accompanying the transitions from it
to the low-symmetry phases are shown in Fig.~\ref{fig2}. The transition to the
$Pb2_1m$ phase is accompanied by out-of-phase displacements of the titanium
and oxygen atoms along the $y$~axis, whereas the cadmium atoms remain almost
unmoved. Upon the transition to the $Pbn2_1$ phase, both metal atoms are
slightly displaced along the $z$~axis (by approximately equal distances),
and the displacements of the titanium and oxygen atoms contain a considerable
$y$-component in addition to the $z$-component.

\begin{table}
\caption{\label{table4}Lattice parameters $a$, $b$, and $c$ (in {\AA}) and
atomic coordinates for ferroelectrically distorted $Pbn2_1$ and $Pb2_1m$ phases
of cadmium titanate.}
\begin{ruledtabular}
\begin{tabular}{ccccc}
          & \multicolumn{2}{c}{$Pbn2_1$} & \multicolumn{2}{c}{$Pb2_1m$} \\
Parameter & This work & Exp.\,(Ref.~\onlinecite{Ferroelectrics.259.85}) & This work & Exp.\,(Ref.~\onlinecite{Ferroelectrics.284.107}) \\
\hline
$a$     & 5.2392     & 5.2946    & 5.2498     & 5.281 \\
$b$     & 5.3777     & 5.4151    & 5.3870     & 5.403 \\
$c$     & 7.6192     & 7.6029    & 7.5699     & 7.583 \\
Cd1$_x$ & $-$0.01101 & $-$0.0083 & $-$0.01400 & $-$0.01106 \\
Cd1$_y$ & +0.04425   & +0.0407   & +0.04583   & +0.04076 \\
Cd1$_z$ & +0.25324   & +0.25     & +0.25000   & +0.25000 \\
Cd2$_x$ & +0.01101   & +0.0083   & +0.00509   & +0.00697 \\
Cd2$_y$ & $-$0.04425 & $-$0.0407 & $-$0.04696 & $-$0.04076 \\
Cd2$_z$ & +0.75324   & 0.75      & $-$0.25000 & $-$0.25000 \\
Ti$_x$  & +0.00080   & +0.004    & +0.00459   & +0.00190 \\
Ti$_y$  & +0.49548   & +0.493    & +0.50699   & +0.5045 \\
Ti$_z$  & +0.00334   & +0.004    & +0.00080   & +0.00214 \\
O1$a_x$ & +0.10277   & +0.091    & +0.10313   & +0.0925 \\
O1$a_y$ & +0.46191   & +0.473    & +0.45559   & +0.4759 \\
O1$a_z$ & +0.24044   & +0.241    & +0.25000   & +0.25000 \\
O1$b_x$ & $-$0.10277 & $-$0.091  & $-$0.09897 & $-$0.0911 \\
O1$b_y$ & $-$0.46191 & $-$0.473  & +0.53144   & +0.5338 \\
O1$b_z$ & +0.74044   & +0.741    & $-$0.25000 & $-$0.25000 \\
O2$a_x$ & +0.68779   & +0.723    & +0.69372   & +0.6999 \\
O2$a_y$ & +0.31805   & +0.308    & +0.29494   & +0.2951 \\
O2$a_z$ & +0.04267   & +0.047    & +0.05436   & +0.0501 \\
O2$b_x$ & +0.29839   & +0.323    & +0.30806   & +0.2989 \\
O2$b_y$ & +0.71471   & +0.710    & +0.68835   & +0.7017 \\
O2$b_z$ & $-$0.06583 & $-$0.045  & $-$0.05264 & $-$0.0457 \\
\end{tabular}
\end{ruledtabular}
\noindent{\footnotesize $^*$ Correct signs of atomic displacements are recovered. \hfill}
\end{table}

The transition to the polar phases results in the appearance of two
nonequivalent sets of oxygen atoms O2 (labeled by the $a$ and $b$ letters in
Table~\ref{table4}), and the disappearance of the $n$ plane upon transition
to the $Pb2_1m$ phase results in the appearance of two nonequivalent cadmium
atoms (Cd1, Cd2). The energy gain (per formula unit) associated with the
distortions is equal to 6.21~meV for the $Pbn2_1$ phase and 1.38~meV for
the $Pb2_1m$ phase.

A comparison of the calculated atomic displacements and the lattice strains
with the results of low-temperature X-ray
measurements~\cite{Ferroelectrics.259.85,Ferroelectrics.284.107}
(Table~\ref{table4}) demonstrates that the atomic displacements and the
lattice strains in the experiment are much smaller. For example, from the
comparison of the calculated lattice parameters in Tables~\ref{table2} and
\ref{table4} it follows
that a considerable increase (by 0.045~{\AA}) in the lattice parameter along
the polar $c$ axis should be observed in the $Pbn2_1$ phase, and the $a$
lattice parameter rather than the $b$ lattice parameter should be increased
in the $Pb2_1m$ phase. In the experiment, the largest spontaneous strain
below the phase transition temperature was observed for the $b$ lattice
parameter (elongation of the order of 0.002~{\AA}),~\cite{Ferroelectrics.259.85}
which disagrees with the predictions for both ferroelectric
phases. The possible causes of these discrepancies will be discussed below.

\subsection{Parameters of an effective Hamiltonian}

Torgashev et al.~\cite{PhysSolidState.47.337} and Gorshunov et
al.~\cite{PhysSolidState.47.547} have supposed that below the
ferroelectric phase transition temperature, one more phase transition
accompanied by a change in the polarization direction occurs in cadmium
titanate. In order to check this hypothesis, we calculated the dependence
of the total energy of the crystal on the lattice distortion with the
$B_{1u}$ and $B_{2u}$ symmetries.

\begin{figure}
\includegraphics[scale=0.75]{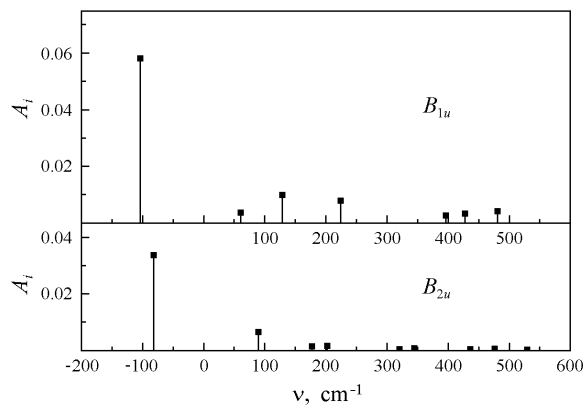}
\caption{Relative contributions of different modes with $B_{1u}$ and $B_{2u}$
symmetries to the ferroelectric distortions accompanying the transitions to
the $Pb2_1m$ and $Pbn2_1$ phases of cadmium titanate.}
\label{fig3}
\end{figure}

As is known, the normal modes form a complete basis set using which any
combination of atomic displacements can be expanded. We expanded the
lattice distortions in the $Pbn2_1$ and $Pb2_1m$ phases determined in
Sec.~\ref{Sec3.2} using the basis of eigenvectors of the dynamic matrix of
CdTiO$_3$ with the $Pbnm$ structure. For the $Pbn2_1$ phase, this expansion
contains the contributions of seven modes with the $B_{1u}$ symmetry (the
acoustic $B_{1u}$ mode that describes a uniform
shift of the unit cell was excluded) and seven fully symmetric
$A_g$ modes (Fig.~\ref{fig3}). For the $Pb2_1m$ phase, the expansion
contains the contributions of nine $B_{2u}$ modes (also without the acoustic
mode) and seven $A_g$ modes. In both cases, the contributions of the
lowest-frequency modes with $B_{1u}$ and $B_{2u}$ symmetries to the total
energy of distortions amount to 92.8 and 95.4\%, respectively. Therefore,
when constructing an effective Hamiltonian, we can restrict ourselves to
the expansion in powers of the amplitudes of these two modes.

The total energy of the crystal was expanded in a Taylor series in powers
of the distortion amplitudes $\xi$ and $\eta$ of the lowest-frequency normal
modes with $B_{1u}$ and $B_{2u}$ symmetries. In this expansion, the lattice
parameters were assumed to be fixed and equal to the lattice parameters of
the $Pbnm$ phase. The resulting expansion had the form
\begin{equation}
E_{\rm tot}(\xi,\eta) = E_{\rm tot}(0,0) + b_1 \xi^2 + c_1 \xi^4 + b_2 \eta^2
+ c_2 \eta^4 + d  \xi^2 \eta^2
\end{equation}
with the coefficients $b_1 = -0.4075$~Ha (Hartree), $b_2 = -0.2614$~Ha,
$c_1 = 183.49$~Ha, $c_2 = 249.58$~Ha, and $d = 457.4$~Ha. It turned out that
the total energy is adequately described by the fourth-order polynomials of
two order parameters and that there is no need to include the sixth-order
invariants into the expansion. This means that the phase transition in
CdTiO$_3$ is far from the tricritical point (the closeness to which was
supposed in Ref.~\onlinecite{PhysSolidState.43.2146}). Furthermore, since we
have $d > 2 \sqrt{c_1 c_2}$, the minima of the total energy correspond to
the order parameters $\pm (\xi, 0)$ and $\pm (0, \eta)$ and are separated
from each other by the energy barriers. This means that the formation of
the monoclinic phase with $Pb$ space group and tilted polarization vector
is energetically unfavorable. The calculation using the formula~(1) shows
that the energy minima are observed at the mode amplitudes of
($\xi = 0.03332$, $\eta = 0$) and ($\xi = 0$, $\eta = 0.02288$) and that the
energy gains associated with these lattice distortions are 6.16 and 1.86~meV,
respectively. They are close to the corresponding energy gains of the true
distortions (see Sec.~\ref{Sec3.2}).

Since only two modes with the lowest frequencies were included in expansion~(1),
it was necessary to check that the energetically most favorable $Pbn2_1$ phase
remains stable with respect to small distortions with the $B_{2u}$ symmetry when
taking into account all modes involved in the distortion. For this purpose,
we calculated the phonon spectrum of CdTiO$_3$ in the $Pbn2_1$ phase. In this
spectrum, the ferroelectric mode polarized along the $y$ axis had the lowest
frequency of 83 cm$^{-1}$. The positive values of all mode frequencies confirm
that the polar $Pbn2_1$ phase is the ground-state structure of CdTiO$_3$.

Although the above analysis of the polar $Pbn2_1$ phase stability is for
$T = 0$, it is unlikely that a change of the $c_1$, $c_2$, and $d$ coefficients
in the thermodynamic potential with increasing temperature can result in the
violation of the $d > 2 \sqrt{c_1 c_2}$ condition and the transition into the
monoclinic phase. Therefore, the possibility of successive ferroelectric phase
transitions in CdTiO$_3$ with varying temperature can be most likely ruled out.
Possibly, the specific features observed in the dielectric constant of CdTiO$_3$
near 50~K~\cite{Ferroelectrics.217.137,PhysSolidState.43.2146,Ferroelectrics.270.381}
are associated with the existence of side minima of the total energy that are
separated from the main minima by low potential barriers.

\subsection{Spontaneous polarization}

The spontaneous polarization $P_s$ in the orthorhombic $Pbn2_1$ and $Pb2_1m$
phases of cadmium titanate was calculated by the Berry's phase
method.~\cite{PhysRevB.66.104108}  The calculated values of $P_s$ corresponding
to the lattice distortions determined in Sec.~\ref{Sec3.2} for the $Pbn2_1$ and
$Pb2_1m$ phases are equal to 0.29 and 0.21~C/m$^2$, respectively. The
spontaneous polarization $P_s$ corresponding to the minima in expansion~(1)
is equal to 0.21~C/m$^2$ for the $Pbn2_1$ phase and 0.16~C/m$^2$ for the
$Pb2_1m$ phase. It is worth noting that both calculated values of $P_s$
considerably exceed the experimentally obtained polarizations of
0.002--0.009~C/m$^2$
(Refs.~\onlinecite{Ferroelectrics.259.85,PhysSolidState.43.2146}).

\section{Discussion}

As shown in Ref.~\onlinecite{PhysSolidState.51.362}, the ferroelectric
instability is characteristic of cubic phases of all ten titanates with the
perovskite structure studied in the cited work. In five of these compounds,
which undergo structural phase transitions to the $Pbnm$ or $I4/mcm$ phases,
the ferroelectric instability is weakened and can be observed only in three
compounds with the $Pbnm$ structure: CdTiO$_3$, ZnTiO$_3$, and MgTiO$_3$
(Table~\ref{table1}). A comparison of the frequencies of the softest
ferroelectric modes at the $\Gamma$~point in the cubic parent phases and in
the $Pbnm$ or $I4/mcm$ phases shows that upon the structural phase transition,
the ferroelectric instability is retained in crystals in which it
is strongest in the cubic phase. Therefore, it is not surprising that the
temperature of the ferroelectric phase transition in CdTiO$_3$ is considerably
lower than in BaTiO$_3$ or PbTiO$_3$.

Studies of SrTiO$_3$ have long established that despite the existence of the
ferroelectric instability in this compound, the corresponding phase transition
does not occur with decreasing temperature. It is believed that such a behavior
is due to quantum fluctuations, i.e., zero-point vibrations of
atoms.~\cite{PhysRevB.19.3593}  One more factor responsible for this behavior
can be the structural phase transition that occurs in this compound, which,
as shown above, weakens the ferroelectric instability.

It is possible that the strong influence of quantum fluctuations on the
physical properties of crystals should also be observed in CdTiO$_3$. As
shown in Ref.~\onlinecite{PhysRevB.53.5047}, quantum effects most strongly
affect the modes with a small ``reduced mass,'' in particular, the
ferroelectric mode to which light oxygen atoms make a significant contribution.
This explains why the substitution of the $^{16}$O isotope by the
heavier $^{18}$O isotope in strontium titanate results in a real ferroelectric
phase transition.~\cite{PhysRevLett.82.3540}

In order to take into account quantum fluctuations, expression (1) for the
``potential'' energy should be complemented by the kinetic energy of nuclei.
When performing first-principles calculations, this energy is ignored in order
to calculate correctly the forces acting on atoms (the Born--Oppenheimer
approximation).

Although the approach used below requires a more rigorous justification, it is
possible to try (following Ref.~\onlinecite{PhysRevB.53.5047}) to take into
account quantum fluctuations at the level of the mode motion by adding the
kinetic energy of the mode described by the $(-\hbar^2/2M^*)\nabla^2$ operator,
where $M^*$ is some reduced mass, to the potential energy~(1). In order to
determine the quantitative criterion of the complete suppression of a phase
transition by quantum fluctuations, we consider a simple one-dimensional
problem in which a particle of mass $M^*$ moves in a double-well potential
$V(x) = -ax^2 + bx^4$. Suppression of the phase transition by quantum
fluctuations occurs at the moment when the kinetic energy of zero-point
mode vibrations becomes higher than the well depth of the potential barrier
under consideration. The numerical solution of the Schr\"odinger equation shows
that this occurs when $b/\sqrt{M^*a^3} > 0.428$. By eliminating the unknown
mass $M^*$ from this condition, the criterion can be rewritten in a physically
clear form $\hbar \nu / E_0 > 2.419$. Here, $\nu = \sqrt{-2a/M^*}/2\pi$ is
the value of an imaginary mode frequency in the high-symmetry configuration
(maximum of the potential barrier) determined in the classical
(Born--Oppenheimer) approximation, and $E_0 = a^2/4b$ is the depth of the
potential well for the $V(x)$ potential. For the four-minimum potential well
described by equation~(1), the quantitative value of the criterion can be
slightly different.

\begin{table}
\caption{\label{table5}Frequencies of unstable ferroelectric modes in
the high-symmetry phases, energy gains resulting from the lattice
distortion,~\cite{PhysSolidState.51.362} and $h\nu /E_0$ ratios for three
ferroelectric phase transitions in SrTiO$_3$ and CdTiO$_3$.}
\begin{ruledtabular}
\begin{tabular}{cccc}
Phase transition & $\nu$ (cm$^{-1}$) & $E_0$ (meV) & $h \nu/E_0$ \\
\hline
SrTiO$_3$, $Pm3m \to R3m$    & 68$i$  & 0.75 & 11.2 \\
CdTiO$_3$, $Pbnm \to Pb2_1m$ & 81$i$  & 1.38 & 7.28 \\
CdTiO$_3$, $Pbnm \to Pbn2_1$ & 104$i$ & 6.21 & 2.08 \\
\end{tabular}
\end{ruledtabular}
\end{table}

We use the obtained criterion to evaluate the degree of influence of quantum
fluctuations on the ferroelectric phase transitions in CdTiO$_3$. The
frequencies of unstable ferroelectric modes determined from first principles
in the Born--Oppenheimer approximation, the energies $E_0$ of the ordered
phases, and the $h\nu/E_0$ ratios for the hypothetical ferroelectric phase
transition in SrTiO$_3$ and two phase transitions in
CdTiO$_3$ are given in Table~\ref{table5}. It follows from the table that
quantum fluctuations should suppress the ferroelectric phase transition in
strontium titanate and the transition to the $Pb2_1m$ phase in cadmium
titanate. As regards the phase transition to the $Pbn2_1$ phase in cadmium
titanate, this transition is susceptible to strong quantum fluctuations but
is not completely suppressed. Therefore, the only phase transition that can
be associated with the experimentally observed ferroelectric phase transition
in CdTiO$_3$ is that to the $Pbn2_1$ phase.

Quantum fluctuations can also be responsible for the significant
discrepancies in the structural positions of atoms and in the values of
the spontaneous polarization $P_s$. In quantum-mechanical calculations of
the ground state for multiwell potentials, the displacement corresponding
to the most probable atomic position is always smaller than the displacement
corresponding to the minimum of the potential energy. Therefore, quantum
fluctuations should decrease the distortions accompanying the phase transition
and decrease the spontaneous polarization $P_s$. Other possible factors
responsible for the decrease in the experimentally obtained polarization in
CdTiO$_3$ can be the manifestation of twinning in
crystals~\cite{PhysSolidState.47.337,ActaCrystC.43.1668,PhysSolidState.42.1329}
and the presence of long-period structures competing with the $Pbnm$
phase.~\cite{PhysSolidState.42.1329}

\section{Conclusions}

The first-principles calculations of the structural parameters and the phonon
spectrum of orthorhombic CdTiO$_3$ enabled to refine the identification
of Raman scattering and IR reflectance spectra. The calculated dependence
of the total energy of cadmium titanate on the amplitudes of two unstable modes
indicates that the ferroelectrically distorted $Pbn2_1$ phase is the ground
state of the crystal at $T = 0$. This phase appears to be the most stable
with respect to quantum fluctuations, which are rather strong and suppress
other possible lattice distortions. Quantum fluctuations were found to be
one of the main factors responsible for the discrepancy between the
calculated and experimentally observed values of spontaneous polarization
and structural distortions accompanying the phase transition.

\begin{acknowledgments}
This work was supported by the Russian Foundation for Basic Research (grant
No. 08-02-01436).
\end{acknowledgments}

\providecommand{\BIBYu}{Yu}

\end{document}